\documentclass[twocolumn,prd,floatfix,nofootinbib]{revtex4}
\usepackage{mathrsfs}
\usepackage{bbm}
\usepackage{amsmath}
\usepackage{graphicx}
\usepackage{dcolumn}
\usepackage{bm}
\usepackage{amssymb}
\usepackage{latexsym}

\def\be{\begin{equation}}
\def\ee{\end{equation}}
\def\ba{\begin{eqnarray}}
\def\ea{\end{eqnarray}}

\begin{document}

\title{Proliferation of the Phoenix Universe
}

\author{Jun Zhang\footnote{Email: junzhang34@gmail.com}}

\affiliation{College of Physical Sciences, Graduate School of
Chinese Academy of Sciences, Beijing 100049, China}

\begin{abstract}

Cyclic cosmology, in which the universe will experience alternating
periods of gravitational collapse and expansion, provides an
interesting understanding of the early universe and is described as
``The Phoenix Universe''. In usual expectation, the cyclic universe
should be homogeneous, however, with studying the cosmological
perturbations, we find that the amplification of curvature
perturbations on the large scale may rip the homogeneous universe
into a fissiparous multiverse after one or several cycles. Thus, we
suggest that the cyclic universe not only rebirths in the ``fire''
and will never ended, like the Phoenix, but also proliferates
eternally.

\end{abstract}

\maketitle

\section{The Cyclic Universe}

As a standard model of our universe, the big bang theory has
succeed in many aspects, especially in explaining the abundance of
primordial elements and prediction of the cosmic microwave
background radiation. However, its successes naturally lead the
physicists to another confused conception: the big bang, which
seems to be the beginning of our universe. Is it really the
beginning of our universe when everything suddenly emerged from
nothing or the bang? It is a radical notion for most physicists.
Actually, in the big bang theory, the big bang is nothing but a
moment when the temperature and density are extremely so high that
the physical theory, as far as we have known, is not available any
more. Then what is the universe before the big bang? The issue is
open.

Historically, an idea is that our universe is always existing with
no beginning and no ending, and the universe we live in today is
just one phase of infinite cyclic periods. Physicists revisited
this idea in the frame of the Einstein Gravitation, and developed a
kind of universe models which are called cyclic universe or
oscillating universe. In these models, the universe is homogeneous
with a periodical scalar factor and will orderly experience the
bounce, expansion, turnaround and contraction in each cycle.
Because the universe in cyclic models always rebirths in ``fire''
and will never end, it is also named as the Phoenix universe.

In different models, the evolution of cyclic universe is driven by
different physics. For example, in \cite{STS}, Steinhardt and Turok
introduce a cyclic model in high dimensional string theory with an
infinite and flat universe, while pointed by Brown\cite{Brown} and
Baum\cite{Baum} etc, the cyclic evolution of the universe can also
be realized with a modified gravitational equations. For other more
cyclic models one can see
\cite{BD,KSS,Piao04,Lidsey04,CB,Xiong,Xin,LS,Biswas,Biswas1,Cai0906},
and also see \cite{NB} for review. Although in some of models the
physics under such high energy is still uncertain, these studies
provide good insights and effective theories to describe the
dynamics of the cyclic universe.

As an alternate model of the early universe, the studies of cyclic
universe model not only bring a distinctive insight for the big
bang described in the standard universe model, but also promote the
study of and succeed in other problems. The explanation of the
primordial perturbations is just one of them.

\section{Cosmological Perturbations in Cycle}

Today, our universe has a well developed nonlinear structure which
takes the form of galaxies, clusters and super-clusters of
galaxies, and of filaments and voids on large scales. However, all
the inhomogeneities in the density distribution, when averaged over
hundreds mega-parsecs, are very small and can be seemed as the
perturbations of the homogeneous background. Moreover, if one
ascend to the time of recombination, it will be found that all of
these nonlinear inhomogeneous structures have not formatted and
only small perturbations, which are called primordial
perturbations, remain. By measuring the cosmic microwave
background, we can detect and even characterize these primordial
perturbations as gaussian, adiabatic fluctuations with a
scale-invariant power spectrum. There have been mature theories of
explaining how these primordial perturbations evolve and finally
lead the structure of our universe today. The details of this
process are so complicate that it can only be calculated
numerically, but fortunately the current supercomputers are strong
enough to deal with these calculations and have given very
beautiful results.

On the other hands, as the initial conditions of structure
formation, the primordial perturbations play an important role in
modern cosmology. However, what is the origin of primordial
perturbations, which is the seed of the structure formation, and
why these perturbations are gaussian, adiabatic and scale-invariant
are key challenges for the modern cosmology.

According to the present understanding, because the physical
wavelength of fixed co-moving scales is increasing less fast than
the Hubble radius in the late times (times later and including the
period of nucleosynthesis) when cosmological evolution is well
described in the big bang theory, we must look to the very early
universe to find an explanation for the observed structures.
Therefore, a successive model of the early universe must give an
appropriate explanation of origin of the primordial perturbations.

Cyclic universe models provide a good explanation about origin of
the primordial perturbations. In cyclic universe models, the prime
perturbations are generated in the previous cycle by quantum
fluctuations. During the contraction, since the Hubble radius
shrink faster than the wavelength of the perturbations with fixed
co-moving scales, the perturbations will leave out of the Hubble
radius. While after the bounce, which is seems like the big bang of
our cycle, the universe expands again, the perturbations leave our
the previous Hubble radius will reenter the horizon and then set as
the seeds of the structure formation.

\section{Proliferation of the Phoenix Universe}

Before we discuss the proliferation of the Phoenix Universe, we
would like to address a little further about the behaviors of the
cosmological perturbations as preparation. Although there are many
cyclic universe models, for the generality of the results, we will
not involve the details of model building. Actually, it is not
difficult, since behaviors of the perturbations is mainly dependent
on the background evolution, namely the evolution of the scale
factor. To be more specifical, we would like to consider the
universe is dominated by matter with equation of state $\omega$,
and the scale factor can be described as $a \sim
\eta^{\frac{n}{n-1}}$ where $\eta$ is the conformal time. If we
solve the equation of the curvature perturbations $\zeta$ in
momentum space, we can find that the linear perturbation with each
$k$ evolves dependently. For the perturbations with wavelength
smaller than Hubble radius, the amplitude of $\zeta_k$ evolves in
inverse proportion to the scale factor. While for perturbations
with wavelength larger than Hubble radius, we can find that the
solution of $k$-mode curvature perturbation $\zeta_k$ can be
written into two terms which can be denote them as $D_1$ and $D_2$
term, and $D_1$ is constant term while $D_2$ is depended on the
background evolution.

Then we can discuss perturbations in the cyclic universe by
considering a cyclic universe model, starting with the turnaround.
The universe is homogeneous with small perturbations generated by
the quantum fluctuations inside the Hubble radius. During the
contraction with $n>{1\over 3}$, some perturbations leave out the
Hubble radius and the amplitude of $\zeta$ will be dominated by
$D_2$ term, which is increased during this time. After the bounce,
the universe comes into the expanding phase in which the $D_2$ term
is decreasing while the $D_1$ term dominates $\zeta_k$. Thus,
perturbations with wavelength large than Hubble radius keep
constant during expanding until they reenter the Hubble radius and
leading the formation of the large structure. In a word, for a
cycle of cyclic universe during the contraction $\zeta_k$ is
increased on super Hubble radius scale, until the bounce of
corresponding cycle, while during the expansion it becomes
constant. Therefore, the net result is that $\zeta_k$ on large
scale is inevitably amplified.

Since the net amplification is different for each $k$, we can find
that for curvature perturbations generated with quantum fluctuation
spectrum, this process will change the spectrum index as
\begin{eqnarray}
n_\zeta-1=3-\left|{3n-1\over n-1}\right|,
\end{eqnarray}
which is scale invariant for $n\simeq {2\over 3}$, i.e. the
contraction with $\omega \simeq 0$. The above estimation shows a
way of how the primordial perturbation with a scale invariant power
spectrum be gained in the cyclic models.

However, the picture may be not that simple as our usual
expectation. As we have concluded, the amplitude of the
perturbations may be amplified during the contraction, and in
\cite{Piao0901,Piao1001,Zhang:2010bb}we suggest that, this
amplification may change the global picture of our cyclic universe.

\begin{figure}
\includegraphics[width=8.5cm]{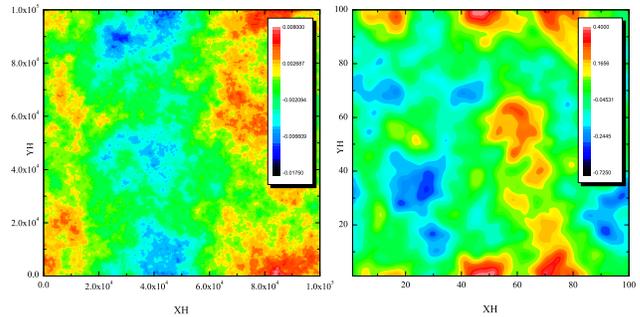}
\caption{\label{fig:4} $\zeta(\vec{x})$ in position space, which
reflects the inhomogeneity of two different backgrounds. The length
scale is in unit of the Hubble radius at that time.}
\end{figure}

We would like to recheck the above process. One should notice that
during the hole process mentioned above, the universe is supposed
to be homogeneous, and evolves with the same background, from the
beginning of contraction to the time of turnaround, which denotes
the end of one cycle, and the net increasing of amplitude of the
perturbations in one cycle, say the $j^{th}$ cycle, can be
described as
\begin{eqnarray}
{\cal P}_{\zeta}^{j+1}(k,\eta_B^{j+1}) \sim k^{\Delta n_s}{\cal
P}_{\zeta}^{j}(k,\eta_B^{j}),
\end{eqnarray}
where $\Delta n_s$ is negative for most of cyclic models. We can
imagine that, due to the amplification, the modes on large scale
will have the amplitude be about order one after many cycles. This
will lead to ${\delta \rho\over \rho} \sim 1$ on corresponding
super horizon scale at this time. Here we use horizon scale, a
different conception with Hubble radius but with same scale in most
of cases, since region outside horizon is causally uncorrelated.

In this case, it is obviously impossible that the different regions
of global universe will evolve synchronously, even if it is
synchronous in previous cycle. Namely, the global universe at the
beginning time of this cycle will be separated into many different
parts, each of which will evolve independently of one another.
While inside any given part, all perturbation modes origin from the
interior of horizon, where is still causally correlative. Thus in
this sense, each of such parts actually corresponds to a new
universe. To illustrate this effect of the perturbation on
background, we show the distribution of curvature perturbation,
which is homogeneous at the very beginning, by transforming the
power spectrum into position space after many cycles in
Fig.\ref{fig:4}.

In principle, each of these new universes will experience the
contraction, bounce and expansion, hereafter all or some of them
will enter into next cycle and proliferate again, and then the
above process may be repeated. In another word, the proliferation
will inevitably occur cycle by cycle, and we can have a cyclic
multiverse scenario. In this cyclic multiverse, the experience of
each universe after proliferation is generally not expected to be
synchronous. Thus when some universe are in a period of matter
domination, it is possible that there are many other universes
which are in the period of contraction or bounce or others. There
is also the proliferation of global universe in chaotic eternal
inflation, in which it is induced by the large quantum fluctuation
of inflation field in its horizon scale. Here, however, the
proliferation is induced by the cyclical amplification of
perturbation on super horizon scale, which is in classical sense,
thus it occurs cycle by cycle.

In conclusion, we have showed that the global configuration of
cyclic universe is more complex than people expected ever, which
actually shows itself a cyclic multiverse. It is found that if the
contracting phase with $w\simeq 0$ is included in each cycle of a
cycle universe, the increasing mode of metric perturbation is
inherited by the constant mode of curvature perturbation that lead
the amplification of metric perturbation on super horizon scale
cycle by cycle. Therefore after few cycles the universe will be
inevitably separated into many parts independent of one another,
each of which corresponds to a new universe and evolve up to next
cycle, and then is separated again. Therefore a cyclic multiverse
scenario, in which the universe proliferates cycle by cycle, is
actually presented. If the cyclic universe is like the Phoenix,
rebirth in the ``fire'' and will never ended, then we suggest that
this Phoenix will also proliferate eternally.

\textbf{Acknowledgments} The author would like to thank the editors
of the Journal of Cosmology for inviting this paper and Prof.
Yun-Song Piao for helpful discussions and comments.

\newpage 


\begin{thebibliography}{99}


\bibitem{STS} P.J. Steinhardt, N. Turok, Science \textbf{296},
(2002) 1436; Phys. Rev. \textbf{D65} 126003 (2002).

\bibitem{Brown}M. G. Brown, K. Freese and W. H. Kinney, JCAP \textbf{0803}, 002 (2008). [9]

\bibitem{Baum}L. Baum and P. H. Frampton, Phys. Rev. Lett. 98, 071301 (2007).


\bibitem{BD} J. Barrow, M. P. Dabrowski, Mon. Not. R. Astr. Soc. \textbf{275}, 850
(1995).

\bibitem{KSS} N. Kanekar, V. Sahni, Y. Shtanov, Phys. Rev. \textbf{D63}, 083520
(2001).

\bibitem{Piao04} Y.S. Piao, Phys. Rev. \textbf{D70}, 101302 (2004);
Y.S. Piao, Y.Z. Zhang, Nucl. Phys. \textbf{B725}, 265 (2005).


\bibitem{Lidsey04} J.E. Lidsey, D.J. Mulryne, N.J. Nunes, R. Tavakol,
Phys. Rev. \textbf{D70}, 063521 (2004).

\bibitem{CB} T. Clifton, J.D. Barrow, Phys. Rev. \textbf{D75}, 043515 (2007).

\bibitem{Xiong} H.H. Xiong, Y.F. Cai, T. Qiu, Y.S. Piao, X.M.
Zhang, Phys. Lett. \textbf{B666}, 212 (2008).


\bibitem{Xin} X. Zhang, Eur. Phys. J. \textbf{C60}, 661 (2009).

\bibitem{LS} J.L. Lehners, P.J. Steinhardt, Phys. Rev. \textbf{D79}, 063503 (2009).


\bibitem{Biswas} T. Biswas, S. Alexander, Phys. Rev. \textbf{D80}, 043511 (2009).

\bibitem{Biswas1} T. Biswas, A. Mazumdar, Phys. Rev. \textbf{D80}, 023519
(2009).

\bibitem{Cai0906} Y.F. Cai, E.N. Saridakis JCAP \textbf{0910}, 020 (2009).

\bibitem{NB} M. Novello, S.E.P. Bergliaffa, Phys. Rept. \textbf{463}, 127 (2008).

\bibitem{Piao0901} Y.S. Piao, Phys. Lett. \textbf{B677}, 1 (2009).

\bibitem{Piao1001} Y.S. Piao, arXiv:1001.0631.


\bibitem{Zhang:2010bb} J.~Zhang, Z.~G.~Liu and Y.~S.~Piao, Phys.\ Rev.\  D {\bf 82}, 123505 (2010) [arXiv:1007.2498 [hep-th]].


\end{thebibliography}
\end{document}